\documentclass{article}
\usepackage{spconf,amsmath,graphicx}

\usepackage{comment}
\usepackage{booktabs}
\usepackage{placeins}
\usepackage{float}
\usepackage[usenames,dvipsnames]{xcolor}
\usepackage[super]{nth}

\title{data2vec-aqc: Search for the right Teaching Assistant in the Teacher-Student training setup}
\name{Vasista Sai Lodagala$^{1\star}$, Sreyan Ghosh$^{2\star}$, S. Umesh$^{1}$
\thanks{\hspace*{-1mm}$^{\star}$These authors contributed equally to this work}}

\address{$^1$Indian Institute of Technology, Madras, India\\  $^2$University of Maryland, College Park, USA}
\begin{document}
%\ninept
%
\maketitle
\begin{abstract}
In this paper, we propose a new Self-Supervised Learning (SSL) algorithm called data2vec-aqc, for speech representation learning from unlabeled speech data. Our goal is to improve SSL for speech in domains where both unlabeled and labeled data are limited. Building on the recently introduced data2vec \cite{baevski2022data2vec}, we introduce additional modules to the data2vec framework that leverage the benefit of data augmentations, quantized representations, and clustering. The interaction between these modules helps solve the cross-contrastive loss as an additional self-supervised objective. data2vec-aqc achieves up to 14.1\%  and 20.9\% relative WER improvement over the existing state-of-the-art data2vec system over the test-clean and test-other sets, respectively of LibriSpeech, without the use of any language model (LM). Our proposed model also achieves up to 17.8\% relative WER gains over the baseline data2vec when fine-tuned on a subset of the Switchboard dataset. 
Code: https://github.com/Speech-Lab-IITM/data2vec-aqc.

%We make all our codes publicly available on GitHub \footnote{We will release codes after paper acceptance}.

%In this paper we propose a new Self-Supervised Learning (SSL) algorithm for speech representation learning from un-labelled speech data. Our goal is to improve SSL for speech in domains where both unlabeled and labeled data is limited. Building on the recently introduced data2vec \cite{baevski2022data2vec}, we make key design choices to make data2vec simultaneously solve an additional contrastive learning task but with key differences from literature including leveraging data augmentations and learning a symmetrical cross-contrastive loss between the student and the teacher networks. data2vec-aqc achieves x\%  and y\% absolute WER improvement for Automatic Speech Recognition (ASR) over the existing state-of-the-art data2vec in our experimental setting on LibriSpeech and SwitchBoard datasets respectively.

% We show that augmentations can improve Masked Acoustic Modelling (MAM) when learning from latent target representations like in data2vec and that learning from quantized targets is an important and necessary step to avoid collapse in trying to solve a contrastive task. 

% Additionally we show the role of augmentation in Masked Acoustic Modelling (MAM) and that learning from quantized targets is an important and necessary step to avoid collapse in trying to solve a contrastive task. 

% we show that data2vec benefits from contrastive loss
\end{abstract}
\begin{keywords}
self-supervised learning, automatic speech recognition, low-resource, domain adaptation
\end{keywords}
\section{Introduction}
\label{sec:intro}
SSL for speech representation learning from unlabeled data has been an active area of research over the past few years \cite{baevski2020wav2vec,hsu2021hubert,chen2022wavlm}. All of these proposed systems try to solve a Masked Acoustic Modeling (MAM) task in some form. data2vec \cite{baevski2022data2vec} was one of the first works to show that learning from latent targets is possible for SSL-based speech representation learning. 
%data2vec solves a MAM-based reconstruction task, wherein the student encoder is fed with a masked version of the audio and the teacher encoder (which is a momentum encoder inspired from \cite{grill2020bootstrap,he2020momentum}) is fed with the complete unmasked version of the audio. Finally, they regress the latent representations from the student and teacher to solve the SSL task.

While it is common knowledge that SSL benefits from scale \cite{hannun2021history}, systems that can learn speech representations with limited amount of unlabeled data is the need of the hour \cite{hannun2021history}. Given that the amount of unlabeled data in most languages is quite limited, SSL methods that can learn meaningful representations even in low-resource regimes (both data and compute) can help in the universal adoption of such systems. Moreover, to date, SSL in speech fails to perform well in instances of domain shift between the unlabeled source and labeled target~\cite{hsu2021robust,sanabria2022measuring}. Data augmentation is often considered to be an effective strategy in the supervised setting, specially in cases of limited access to labeled data \cite{ko2015audio}.
Previous to its adoption in Speech as MAM, SSL in Computer Vision (CV) existed as a task that focuses on learning to identify randomly augmented versions of the same image. %Inspired from this, recent works in speech \cite{sriram2022wav2vec,lodagala2022ccc} try to introduce augmentation in MAM, wherein the model now tries to identify the true quantized representation from an augmented version of the same audio sample.
Inspired by this, recent works in speech \cite{sriram2022wav2vec,lodagala2022ccc} exploit data augmentations to improve the performance and generalizability of SSL models.
We adopt this simple yet powerful idea to learn useful speech representations efficiently when unlabeled data for a domain is scarce.

{\noindent \bf Main Contributions:} In this paper, we propose data2vec-aqc, a novel SSL-based pre-training methodology for learning speech representations from low-resource unlabeled speech. We build on data2vec and achieve this by proposing several improvements to it. First, we make data2vec simultaneously solve a MAM-based cross-contrastive task between the student and teacher networks by passing randomly augmented version(s) of the same audio sample passed through each network.
We add a quantizer module similar to \cite{baevski2020wav2vec}, as sampling negatives from the quantized representations has been proven to be effective.
%Precisely, we add a quantizer module similar to \cite{baevski2020wav2vec} and allow the output of the teacher to find it's true quantized representation from the feature extractor representation of the augmented audio originally input to the student and vice-versa. 
Additionally, we introduce a clustering module \cite{lodagala2022ccc}, to cluster the quantized representations and control the effect of those negatives in the contrastive loss computation, that share the same cluster as the positive. 
%Intuitively, this helps the system to contrast against more informative negatives. 
Our proposed data2vec-aqc achieves significant improvements over the standard data2vec framework when working with a limited amount of pre-training data. Additionally, when pre-trained on large-scale unlabeled speech (960 hours of unlabeled LibriSpeech data), our model performs significantly better than the baseline data2vec model over the several downstream tasks presented over SUPERB\footnote{https://superbbenchmark.org/leaderboard?subset=Public+Set} \cite{yang2021superb}. 

% First we prepare 2 randomly augmented versions of the same audio sample and pass one through the student while the other through the teacher network. Next, similar to prior work \cite{baevski2022data2vec,baevski2020wav2vec}, post the convolutional feature extraction stage, we randomly mask $n\%$ of the samples for input to the student transformer encoder but keep representations in teacher unmasked. Then we add a quantizer similar to \cite{baevski2020wav2vec} and pass the feature extractor outputs for both the student and the teacher to the quantizer. Post this we try to solve a symmetric cross-contrastive task whereby for each masked masked timestep from the student, we try to identify the true quantized representation from the other augmentation originally passed to the teacher network and vice-versa. Additionally, for each contrastive loss calculation, we cluster the quantized representations using kmeans and diminish the effect of those negatives in the loss computation that fall into the same cluster as the positive. Intuitively, this helps the system to contrast against more informative negatives for the contrastive loss. In the next few sections, we describe in detail each added module and our rationale behind adding it.
% and to maintain symmetricity between the student and the teacher, we 

\begin{figure*}[t]
\centering
\includegraphics[width=\textwidth]{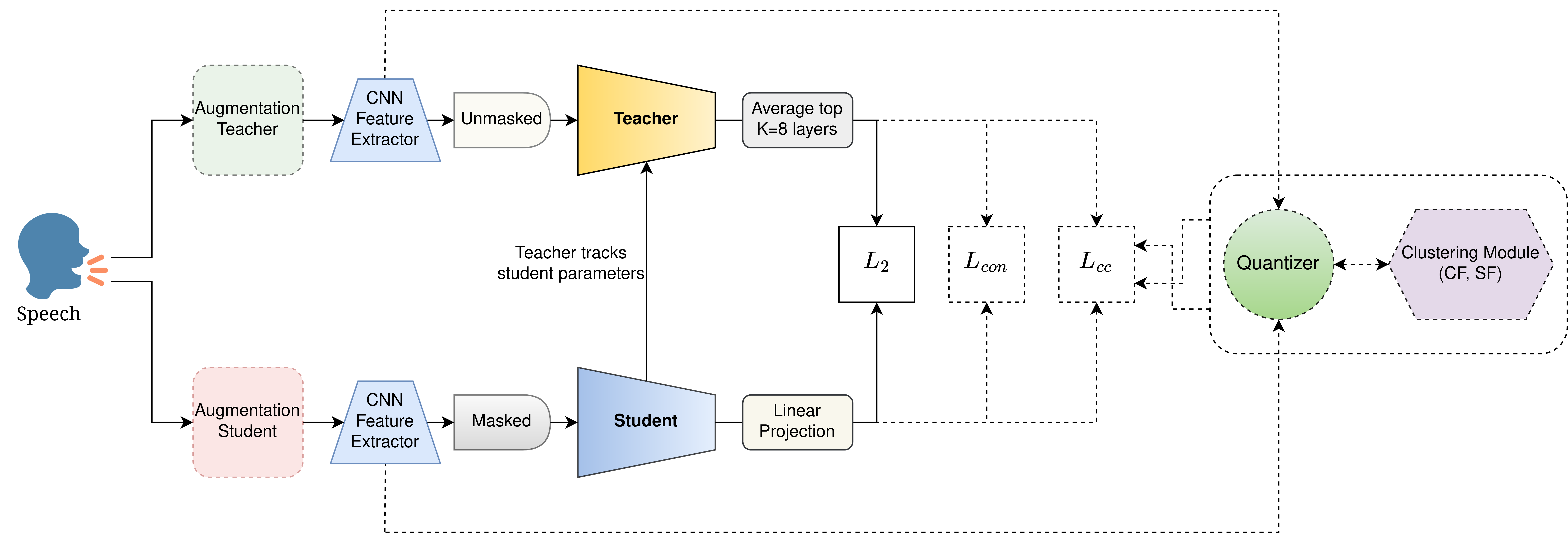}
\caption{\small Illustration of the data2vec-aqc framework. Solid lines represent the original data2vec \cite{baevski2022data2vec} SSL process wherein the student is fed with the masked version of the audio sample while the teacher gets to see the unmasked version. Finally, an $L_2$-loss is computed between the outputs of the two networks. The dashed lines and entities represent our added modules to data2vec which enables it to calculate $L_{con}$ and $L_{cc}$. We elaborate on each added component in Section \ref{sec:methodlogy}.}
\label{fig:data2vec-aqc}
\end{figure*}

\section{Methodology}
\label{sec:methodlogy}

The standard data2vec architecture involves a student and teacher network, both of which see raw speech as the input, and the teacher's parameters are updated based on an exponential moving average of the student's (a momentum encoder inspired from \cite{grill2020bootstrap,he2020momentum}). A simple $L_2$-loss is computed between the student embedding (after a linear projection) and the average of the embeddings from the top 8 layers of the teacher network. Though there is an option to switch to $L_1$-loss in the data2vec setup, the authors of \cite{baevski2022data2vec} find that a simple $L_2$-loss works well for speech processing. Following the solid lines and entities in solid borders in Fig.\ref{fig:data2vec-aqc} would illustrate the standard data2vec architecture described above.

While the data2vec framework has shown remarkable results across a variety of SLP tasks \cite{yang2021superb}, in this paper, we take a step further and focus on finding the right ``teaching assistant" for the existing data2vec student-teacher learning framework. To achieve this, we introduce 3 additional components to the existing data2vec setup, which are: \textbf{a}ugmentation(s), a \textbf{q}uantizer module, and a \textbf{c}lustering module. We call our approach data2vec-aqc and highlight these components in the dashed borders of Fig.\ref{fig:data2vec-aqc}. The following subsections elaborate on each of these components, and how each of them independently and jointly leads to the success of data2vec-aqc.

% have been integrated into the standard data2vec framework, and also describe their interaction with each other. 

% which form the ``aqc" of data2vec-aqc are the ``teaching assistants" of our interest. These are essentially the entities in the dashed borders of Fig.\ref{fig:data2vec-aqc}.
% As mentioned in Fig.\ref{fig:data2vec-aqc}, these entities are added to the standard data2vec setup, while retaining the structure of the original setup.

\vspace{-0.5em}
\subsection{Augmentations (data2vec-a)}
%While the standard data2vec architecture extracts features from the raw speech which would then be passed to the student and teacher networks, we can choose to apply augmentations over the original speech samples before features are extracted from them.
The primary focus of data2vec-a is to apply augmentation(s) to the raw audio in the standard data2vec framework before the feature extraction stage. Table \ref{tab:aug} demonstrates the effect of the different augmentations applied. The augmentation choices and their effects have been described in Section \ref{sec:results}. Adding this component to the data2vec setup leads to no change in the loss computation and we still use the standard data2vec loss given by, $L_2 = \frac{1}{2} (s_t - y_t)^2$,
% \begin{equation}
% \label{eqn:mse}
%     L_2 = \frac{1}{2} (s_t - y_t)^2
% \end{equation}
where $s_t$ represents the embedding from the student network for the masked time step $t$ and $y_t$ represents the corresponding embedding from the average of the top-8 layers of the teacher network.

While the standard data2vec-a just solves just the $L_2$-loss, we attempt to introduce additional losses and study their effects. Owing to its immense success in SSL-based speech representation learning \cite{baevski2020wav2vec,jiang2020speech,oord2018representation}, we choose to solve an additional contrastive loss between the latent embeddings of the student and the teacher. Precisely, given an audio sample $X$, let $S_t$ represent the student embeddings, and $Y_t$ represent the embeddings from the teacher over all the masked time steps. By sampling the negatives from the teacher embeddings, we define the contrastive loss over each masked time-step $t$ as:
\begin{equation}
\label{eqn:contr}
    L_{con} = - log \frac{exp(sim(s_t, y_t)/\kappa)}{\sum_{\tilde{y} \sim Y_t} exp(sim(s_t, \tilde{y})/\kappa)}
\end{equation}
where $s_t \in S_t$, $y_t \in Y_t$ and $sim(\mathbf{a},\mathbf{b}) = \mathbf{a}^T\mathbf{b} / \Vert \mathbf{a} \Vert  \Vert \mathbf{b} \Vert$ computes the cosine similarity between the student and teacher representations.
The temperature parameter is represented by $\kappa$.
For the rest of the paper, we denote the standard data2vec with augmentation(s) as data2vec-a, the standard network trained with $L_2 + L_{con}$ as data2vec $+ L_{con}$ and finally, the standard network trained with $L_2 + L_{con}$ and input augmentation(s) as data2vec-a $+ L_{con}$.

% data2vec $+ L_{con}$ represents the network trained with $L_2 + L_{con}$ as the loss, and finally, data2vec-a $+ L_{con}$ would indicate a network trained with $L_2 + L_{con}$ as the loss, along with the augmentation(s) of input(s).

%However, we have a choice to apply augmentations to the inputs of both student and teacher or restrict and apply the augmentation

\vspace{-0.5em}
\subsection{Quantized Representations (data2vec-aq)}
Sampling positive and negative examples from discrete quantized representations for contrastive loss calculation on speech representations have proven to be effective in \cite{baevski2020wav2vec} over the originally proposed \cite{chen2020simple} which calculates the same over latent network representations. The quantizer module integrated into data2vec-aq borrows its design from \cite{baevski2020wav2vec}. Formally put, let $X^s$ and $X^y$ represent the features extracted from the inputs to the student and teacher, respectively. Passing these embeddings through the quantizer yields the discrete representations $Q^s$ and $Q^y$, respectively.
Findings from \cite{lodagala2022ccc} establish the benefit of using a cross-contrastive loss when using augmentations. As data2vec-aq makes use of augmentation(s), we now plugin the cross-contrastive loss $L_{cc}$ to the data2vec framework. Thus, if $S_t$ represents the student embeddings and $Y_t$ represents the embeddings from the teacher, over all the masked time-steps, we define the following loss terms over each masked time-step $t$ as:
\begin{equation}
\label{eqn:s-cross}
    L_{s-cross} = - log \frac{exp(sim(s_t, q_t^{y})/\kappa)}{\sum_{\tilde{q} \sim Q_t^{y}} exp(sim(s_t, \tilde{q})/\kappa)}
\end{equation}
\begin{equation}
\label{eqn:t-cross}
    L_{t-cross} = - log \frac{exp(sim(y_t, q_t^{s})/\kappa)}{\sum_{\tilde{q} \sim Q_t^{s}} exp(sim(y_t, \tilde{q})/\kappa)}
\end{equation}
where, $s_t \in S_t$, $y_t \in Y_t$. Equations \ref{eqn:s-cross} and \ref{eqn:t-cross} compute the contrastive loss between the student embeddings and the quantized representations of the teacher's input and vice-versa for the masked time steps.
We now define the overall cross-contrastive loss $L_{cc}$ as, $L_{cc} = \alpha L_{s-cross} + \beta L_{t-cross}$.
% \begin{equation}
% \label{eqn:cc}
%     L_{cc} = \alpha L_{s-cross} + \beta L_{t-cross}
% \end{equation}
%where, $\alpha$ and $\beta$ are the scalar hyper-parameters that would determine the importance of each of the individual loss terms. 
The overall loss function for data2vec-aq is $L_2 + L_{cc}$, with $\alpha = \beta = 0.5$. The diversity loss from \cite{baevski2020wav2vec} is also a part of the final loss computation as we have a quantizer in place.

\vspace{-0.5em}
\subsection{Clustering of Negatives (data2vec-aqc)}
As suggested by \cite{lodagala2022ccc}, using a k-means clustering module to segregate negative examples and controlling the effect of weak non-informative negative examples helps the overall contrastive learning task. We make use of the same clustering module \cite{lodagala2022ccc} which has cluster factor $(CF)$ and scale factor $(SF)$ as its hyper-parameters. If $NF$ is the number of frames per speech sample in a mini-batch of audios, then the number of clusters per each audio in this mini-batch would be $ceil(NF/CF)$. 
Upon clustering $Q_t$, we identify the cluster to which the positive sample $q_t$ belongs. If $q_t$ belongs to the cluster $R$, our objective would then be to control the ``influence" of those negatives that share the same cluster $R$. As suggested in \cite{lodagala2022ccc}, we scale down the cosine similarity values of the negative examples from $R$ with the anchor $c_t$, by a scaling factor $SF$. Let $Q^{*}$ denote the sampled set of negative examples. i.e., $Q^{*} = \{ \tilde{q} \sim Q_t \}$ and let the samples in $Q^{*}$ be represented by $q$. The formula for contrastive loss with an integrated clustering module would then be:
\begin{equation}
\label{eqn:clus_contr}
    L_c = - log \frac{e^{(sim(c_t, q_t)/\kappa)}}{\displaystyle\sum_{q \in R} e^{(sim(c_t, q) . SF/\kappa)} + \displaystyle\sum_{q \not \in R} e^{(sim(c_t, q)/\kappa)}}
\end{equation}

In equation \ref{eqn:clus_contr} we demostrate that the influence of negative examples sharing the same cluster as the positive is guided by the scalar $SF$. The overall loss function for data2vec-aqc would still be $L_2 + L_{cc}$, but with each of the contrastive loss terms in $L_{cc}$ taking the form of equation \ref{eqn:clus_contr}. We choose $CF=16$ and $SF=0.3$  in the pooled setting as the hyper-parameters for the clustering module after observing the results in \cite{lodagala2022ccc}.

\section{Experimental Setup}
\label{sec:exp_setup}
\vspace{-0.5em}
\subsection{Pre-training}
In Tables \ref{tab:aug} and \ref{tab:aqc}, we present results for the data2vec\textsubscript{BASE} architecture (12 layers) pre-trained over the 360-hour split of the LibriSpeech dataset \cite{panayotov2015librispeech}.
%All models were pre-trained using the fairseq toolkit \cite{ott2019fairseq}.
%Given that our focus is on developing SSL models with limited amounts of unlabeled data, we have based our experiments on the 360-hour split.
We have based our experiments on the 360-hour split, with a focus on developing SSL models using limited amounts of unlabeled data.
%Owing to the compute resource constraints, the entire 960 hours couldn't be used for pre-training. Upon pre-training our model on LibriSpeech-960h, we would evaluate the same over the SUPERB benchmark \cite{yang2021superb}. 
All models have been pre-trained for 88750 updates (or 250 epochs) over LibriSpeech-360h on 4 A-100 GPUs, with the maximum number of tokens per GPU being 3.8 million. All other hyper-parameters were borrowed from the original setting in \cite{baevski2022data2vec, ott2019fairseq}.
We evaluate the proposed data2vec-aqc BASE model pre-trained on 960h of LibriSpeech, over the array of downstream tasks presented by SUPERB \cite{yang2021superb}.

% Rest of the parameters follow standard configurations made available through \cite{baevski2020wav2vec, ott2019fairseq}.

%Owing to the compute resource constraints, the entire 960 hours couldn't be used for pre-training.
\vspace{-1.0em}
\subsection{Fine-tuning}

% Fine-tuning the pre-trained data2vec models is performed by adding a randomly initialized output layer on top of the Transformer to predict characters \cite{baevski2020wav2vec}. CTC loss has been used during the fine-tuning stage. The plugins added to the standard data2vec framework do not feature during the fine-tuning stage. The feature extractor, the student network and the output layer are the only entities in the fine-tuning stage for all of the models including the baseline.

For fine-tuning the pre-trained data2vec models, we drop all the added pre-training modules and just add a linear output layer on top of the encoder stack to solve the CTC task \cite{baevski2020wav2vec}. 
%The plugins added to the standard data2vec framework do not feature during the fine-tuning stage. 
The pre-trained models presented in Tables \ref{tab:aug} and \ref{tab:aqc} were fine-tuned for 36400 updates on the 100h split of the LibriSpeech dataset.
To demonstrate the robustness of data2vec-aqc's pre-training approach, we also fine-tune these pre-trained models for 11000 updates, on the 30-hour split of the Switchboard dataset \cite{225858}. Switchboard data being telephonic and conversational, these results showcase the generalization capabilities of the proposed approach.

\vspace{-0.5em}
\begin{table}[!h]
\small
% \setlength{\tabcolsep}{8pt}
% \centering
\begin{center}
  \caption{Effect of different augmentations and $L_{con}$}

  \label{tab:aug}
 
  \begin{tabular}{l l l l l l}
    \toprule
    Model & & \multicolumn{2}{c}{dev} & \multicolumn{2}{c}{test}\\
    \cline{3-4} \cline{5-6}
    & & clean & other & clean & other\\ 
    
    % \textbf{Baseline} & & & & & & & & & & & & \\
    \toprule
    %  & & & & & \\
    Baseline data2vec & & 6.4 & 17.5 & 6.4 & 17.7\\
     & & & & & \\
    data2vec + $L_{con}$ & & 9.3 & 23.4 & 9.6 & 23.9\\
    data2vec-a (I) & & 6.4 & 16.5 & 6.6 & 16.8\\
    data2vec-a (II) & & \bf6.1 & \bf15.5 & \bf6.2 & \bf16.0\\
    %Augmentation III & & \bf5.8 & 15.5 & \bf6.1 & 15.6\\
    %Dual Augmentation & & 6.1 & \bf15.2 & 6.2 & \bf15.4\\
    data2vec-a + $L_{con}$ & & 6.6 & 17.2 & 6.8 & 17.5\\
    
    \bottomrule
    
  \end{tabular}
  
\end{center}
\end{table}

\vspace{-2.0em}
\section{Results and Analysis}
\label{sec:results}

\begin{table*}
\small
\setlength{\tabcolsep}{6.6pt}
% \centering
\begin{center}
  \caption{data2vec-aqc performance (\% WER) over different test sets without use of a language model.}

  \label{tab:aqc}

  \begin{tabular}{l l l l l l l l l l l l l}
    \toprule
    Model & & \multicolumn{2}{c}{dev} & & \multicolumn{2}{c}{test} & & \multicolumn{2}{c}{WSJ} & & Switchboard\\
    \cline{3-4} \cline{6-7} \cline{9-10}
    & & clean & other & & clean & other & & dev93 & eval93 & & Dev\\ 
    
    % \textbf{Baseline} & & & & & & & & & & & & \\
    \toprule
    %  & & & & & & & & & & & & \\
    
    \textbf{960h Pretraining for 400K updates} & & & & & & & & & & & & \\
    data2vec BASE \cite{baevski2022data2vec} & & 4.2 & 9.6 & & 4.2 & 9.7 & & 20.4 & 20.0 & & \\
    
    wav2vec 2.0 BASE \cite{baevski2020wav2vec} & & 6.1 & 13.8 & & 6.1 & 13.5 & & 22.8 & 22.3 & & \\
    
    & & & & & & & & & & & & \\
    
    \textbf{360h Pretraining for 88750 updates} & & & & & & & & & & & & \\
    Baseline data2vec & & 6.4 & 17.5 & & 6.4 & 17.7 & & 23.1 & 22.3 & & 21.3\\
    
    data2vec-a & & 6.1 & 15.5 & & 6.2 & 16.0 & & 23.1 & 22.8 & & 18.9\\
    
    data2vec-aq & & 5.7 & 15.0 & & 6.0 & 15.2 & & 22.7 & 22.7 & & 18.2\\
    
    data2vec-aq (Dual Augmentation) & & 7.0 & 17.7 & & 7.1 & 18.4 & & 24.6 & 23.7 & & 21.4\\
    
    data2vec-aqc & & \bf5.3 & \bf13.9 & & \bf5.5 & \bf14.0 & & \bf22.0 & \bf21.7 & & \bf17.5\\
    
    \bottomrule
    
  \end{tabular}
  
\end{center}
\end{table*}

Results from Tables \ref{tab:aug} and \ref{tab:aqc}, are without the use of any LM.
\vspace{1mm}

\noindent \textbf{data2vec-a and $L_{con}$}: Results from Table \ref{tab:aug} indicate that adding an additional contrastive loss, degrades the performance of data2vec. However, introducing augmentations to the data2vec framework helps improve the performance over the baseline. The augmentation applied in data2vec-a (II) is an amalgamation of 3 different augmentation strategies. The audio sample is augmented with additive noise at a random signal-to-noise ratio (SNR) between 3dB and 15dB, with a probability of $0.6$. With a probability of $0.7$, the speech sample is the convolved with a random Reverberation Impulse Response (RIR). Eventually, with a probability of $0.8$, at a signal-to-noise ratio (SNR) between 0dB and 15dB, background noise has been added from random noise samples of the noise set from the MUSAN corpus \cite{musan2015}. The reason behind this specific set of augmentations for data2vec-a (II) arises from the work of \cite{https://doi.org/10.48550/arxiv.2010.12715}, which uses augmentations in a supervised setting and \cite{lodagala2022ccc}, which makes use of the same in a self-supervised setting. data2vec-a (I) on the other hand, makes use of the same set of augmentations but with no probability associated with any of the augmentations involved. In other words, additive noise, reverberation, and random noise samples are always applied in the case of data2vec-a (I). It is to be noted that, in the case of data2vec-a (I) and data2vec-a (II), the augmentations are applied only to the input to the student, and the teacher network gets the original sample as its input. Since data2vec-a (II) has the better performing augmentation, for ease of reference, when we refer to data2vec-a further, it indicates data2vec-a (II). We notice in data2vec-a $+ L_{con}$, that the gains obtained from data2vec-a are offset when we add an additional contrastive loss. This again indicates that sampling negatives from latent representations might not be the best choice when implementing a contrastive loss and provides the motivation to integrate a quantization module.

\vspace{-0.5em}
\noindent \textbf{data2vec-aq and data2vec-aqc}: From the results in Table \ref{tab:aqc}, we notice that with the help of a quantization module and a cross-contrastive loss ($L_{cc}$), data2vec-aq outperforms data2vec-a. This re-emphasizes the need to sample negative examples from quantized representations and also demonstrates the effectiveness of the cross-contrastive loss. To observe the effect of passing an augmented input to the teacher network, we pass the augmentation from data2vec-a (I) to the teacher and the augmentation from data2vec-a (II) to the student. However, we notice that data2vec-aq (Dual Augmentation) under-performs data2vec-a. Results from \cite{ressl} suggest that using a weak-augmentation strategy for teacher is beneficial. The augmentation from data2vec-a (I) not being a weak augmentation explains this observation.
Gains from efficient negative sampling through a clustering module can be observed with data2vec-aqc outperforming data2vec-aq. The proposed data2vec-aqc achieves upto 14.1\% and 20.9\% relative WER improvement compared to the Baseline data2vec over the test-clean and test-other sets, respectively of LibriSpeech, when fine-tuned on LibriSpeech-100h split.
% \vspace{-0.5em}

\noindent \textbf{Comparison with the 960h pre-trained models}: Table \ref{tab:aqc} also presents the results of wav2vec 2.0 BASE and data2vec BASE that have been pre-trained on LibriSpeech-960h for 400K updates and fine-tuned on LibriSpeech-100h for 80K updates. These fine-tuned models have been sourced from the corresponding repositories of fairseq \cite{ott2019fairseq}. Though data2vec-aqc has been pre-trained only on the 360h split for 88750 updates and fine-tuned on LibriSpeech-100h for 36400 updates, it competes in terms of performance with wav2vec 2.0 BASE. These results have been presented to demonstrate the effectiveness of data2vec-aqc's pre-training approach.
% \vspace{-0.5em}

\noindent \textbf{Adaptation and SUPERB}:
data2vec-aqc model fine-tuned on LibriSpeech-100h outperforms the baseline data2vec model over the dev93 and eval93 sets of the WSJ dataset \cite{paul1992design}. It is to be noted that the data from the WSJ dataset was not used in either the pre-training or the fine-tuning stages.
When fine-tuned on the 30-hour subset of the Switchboard data, datavec-aqc outperforms the baseline data2vec model by 17.8\% relative WER, thereby demonstrating its domain adaptation capabilities.
data2vec-aqc BASE pre-trained on LibriSpeech-960h has been evaluated over SUPERB \cite{yang2021superb} and is ranked \nth{5} over the Challenge public set, significantly outperforming the datavec BASE model which is ranked \nth{13} in the same leaderboard.
Also, data2vec-aqc is among the best performing models for the Speech Enhancement task.
%data2vec-aqc fine-tuned on LibriSpeech-100h achieves up to 4.7\% and 2.7\% relative WER improvement compared to the Baseline data2vec over the dev93 and eval93 sets of the WSJ dataset \cite{paul1992design}. This shows the efficiency of our pre-training approach over unseen data. 
%Also, the robustness of data2vec-aq is evident from its domain adaptation to the Switvhboard data as it achieves upto 12.7\% absolute WER improvement compared to data2vec over the Switchboard dev set.
%Also, the robustness of the pre-training approach is evident from its domain adaptation to the Switchboard data. When data2vec-aqc is fine-tuned on the 30-hour subset of the Switchboard data, the model achieves up to 17.8\% relative WER improvement compared to Baseline data2vec over the Switchboard dev set.
\vspace{-0.75em}
\section{Conclusion}

In this paper, we present data2vec-aqc, a novel SSL-based pre-training approach based on data2vec that improves speech representation learning with limited amounts of unlabeled data. As a part of our future work, we would like to explore and enhance data2vec-aqc's performance on various other downstream tasks.
%In this paper, we present data2vec-aqc, a novel SSL-based pre-training approach based on data2vec that improves speech representation learning on the original data2vec with limited amount of unlabeled data. As a part of our future work, we would like to explore and enhance data2vec-aqc's performance on various other downstream tasks.
%We evaluate our learned representations on the task of ASR and show absolute WER improvements of 5.3\% and 12.9\% on LibriSpeech test clean and test other sets and 12.7\% absolute WER improvement SWBD over the existing SOTA data2vec. 

%In this paper, we introduce ccc-wav2vec 2.0, a novel SSL-based pre-training approach based on wav2vec 2.0, that improves on the original wav2vec 2.0 discrimination task. Our approach consistently outperforms the wav2vec 2.0 architecture while training the same number of parameters. As a part of future work, we would like to explore techniques to sample negatives that make the instance discrimination task harder to solve and also explore new data augmentation techniques.

\vfill\pagebreak

% \section{REFERENCES}
% \label{sec:refs}

%List and number all bibliographical references at the end of the paper. The references can be numbered in alphabetic order or in order of appearance in the document. When referring to them in the text, type the corresponding reference number in square brackets as shown at the end of this sentence \cite{C2}. An additional final page (the fifth page, in most cases) is allowed, but must contain only references to the prior literature.

\bibliographystyle{IEEEbib}
\bibliography{strings,refs}

\begin{thebibliography}{10}

\bibitem{baevski2022data2vec}
Alexei Baevski, Wei-Ning Hsu, Qiantong Xu, Arun Babu, Jiatao Gu, and Michael
  Auli,
\newblock ``Data2vec: A general framework for self-supervised learning in
  speech, vision and language,''
\newblock {\em arXiv preprint arXiv:2202.03555}, 2022.

\bibitem{baevski2020wav2vec}
Alexei Baevski, Yuhao Zhou, Abdelrahman Mohamed, and Michael Auli,
\newblock ``wav2vec 2.0: A framework for self-supervised learning of speech
  representations,''
\newblock {\em NeurIPS 2020}, pp. 12449--12460.

\bibitem{hsu2021hubert}
Wei-Ning Hsu, Benjamin Bolte, Yao-Hung~Hubert Tsai, Kushal Lakhotia, Ruslan
  Salakhutdinov, and Abdelrahman Mohamed,
\newblock ``Hubert: Self-supervised speech representation learning by masked
  prediction of hidden units,''
\newblock {\em IEEE/ACM Transactions on Audio, Speech, and Language
  Processing}, vol. 29, pp. 3451--3460, 2021.

\bibitem{chen2022wavlm}
Sanyuan Chen, Chengyi Wang, Zhengyang Chen, Yu~Wu, Shujie Liu, Zhuo Chen, Jinyu
  Li, Naoyuki Kanda, Takuya Yoshioka, Xiong Xiao, et~al.,
\newblock ``Wavlm: Large-scale self-supervised pre-training for full stack
  speech processing,''
\newblock {\em IEEE Journal of Selected Topics in Signal Processing}, 2022.

\bibitem{hannun2021history}
Awni Hannun,
\newblock ``The history of speech recognition to the year 2030,''
\newblock {\em arXiv preprint arXiv:2108.00084}, 2021.

\bibitem{hsu2021robust}
Wei-Ning Hsu, Anuroop Sriram, Alexei Baevski, Tatiana Likhomanenko, Qiantong
  Xu, Vineel Pratap, Jacob Kahn, Ann Lee, Ronan Collobert, Gabriel Synnaeve,
  et~al.,
\newblock ``Robust wav2vec 2.0: Analyzing domain shift in self-supervised
  pre-training,''
\newblock {\em arXiv preprint arXiv:2104.01027}, 2021.

\bibitem{sanabria2022measuring}
Ramon Sanabria, Wei-Ning Hsu, Alexei Baevski, and Michael Auli,
\newblock ``Measuring the impact of individual domain factors in
  self-supervised pre-training,''
\newblock {\em arXiv preprint arXiv:2203.00648}, 2022.

\bibitem{ko2015audio}
Tom Ko, Vijayaditya Peddinti, Daniel Povey, and Sanjeev Khudanpur,
\newblock ``Audio augmentation for speech recognition,''
\newblock in {\em 16th annual conference of the ISCA 2015}.

\bibitem{sriram2022wav2vec}
Anuroop Sriram, Michael Auli, and Alexei Baevski,
\newblock ``Wav2vec-aug: Improved self-supervised training with limited data,''
\newblock {\em arXiv preprint arXiv:2206.13654}, 2022.

\bibitem{lodagala2022ccc}
Vasista~Sai Lodagala, Sreyan Ghosh, and S~Umesh,
\newblock ``Ccc-wav2vec 2.0: Clustering aided cross contrastive self-supervised
  learning of speech representations,''
\newblock {\em arXiv preprint arXiv:2210.02592}, 2022.

\bibitem{yang2021superb}
Shu-wen Yang, Po-Han Chi, Yung-Sung Chuang, Cheng-I~Jeff Lai, Kushal Lakhotia,
  Yist~Y Lin, Andy~T Liu, Jiatong Shi, Xuankai Chang, Guan-Ting Lin, et~al.,
\newblock ``Superb: Speech processing universal performance benchmark,''
\newblock {\em arXiv preprint arXiv:2105.01051}, 2021.

\bibitem{grill2020bootstrap}
Grill et~al.,
\newblock ``Bootstrap your own latent-a new approach to self-supervised
  learning,''
\newblock {\em NeurIPS 2020}, vol. 33, pp. 21271--21284.

\bibitem{he2020momentum}
Kaiming He, Haoqi Fan, Yuxin Wu, Saining Xie, and Ross Girshick,
\newblock ``Momentum contrast for unsupervised visual representation
  learning,''
\newblock in {\em Proceedings of the IEEE/CVF conference on computer vision and
  pattern recognition}, 2020, pp. 9729--9738.

\bibitem{jiang2020speech}
Jiang et~al.,
\newblock ``Speech simclr: Combining contrastive and reconstruction objective
  for self-supervised speech representation learning,''
\newblock {\em arXiv preprint arXiv:2010.13991}, 2020.

\bibitem{oord2018representation}
Aaron van~den Oord, Yazhe Li, and Oriol Vinyals,
\newblock ``Representation learning with contrastive predictive coding,''
\newblock {\em arXiv preprint arXiv:1807.03748}, 2018.

\bibitem{chen2020simple}
Ting Chen, Simon Kornblith, Mohammad Norouzi, and Geoffrey Hinton,
\newblock ``A simple framework for contrastive learning of visual
  representations,''
\newblock in {\em ICML 2020}, pp. 1597--1607.

\bibitem{panayotov2015librispeech}
Vassil Panayotov, Guoguo Chen, Daniel Povey, and Sanjeev Khudanpur,
\newblock ``Librispeech: an asr corpus based on public domain audio books,''
\newblock in {\em IEEE ICASSP 2015}, pp. 5206--5210.

\bibitem{ott2019fairseq}
Myle Ott, Sergey Edunov, Alexei Baevski, Angela Fan, Sam Gross, Nathan Ng,
  David Grangier, and Michael Auli,
\newblock ``fairseq: A fast, extensible toolkit for sequence modeling,''
\newblock in {\em NAACL-HLT 2019: Demonstrations}.

\bibitem{225858}
J.J. Godfrey, E.C. Holliman, and J.~McDaniel,
\newblock ``Switchboard: telephone speech corpus for research and
  development,''
\newblock in {\em ICASSP 1992}.

\bibitem{musan2015}
David Snyder, Guoguo Chen, and Daniel Povey,
\newblock ``{MUSAN}: {A} {M}usic, {S}peech, and {N}oise {C}orpus,'' 2015,
\newblock arXiv:1510.08484v1.

\bibitem{https://doi.org/10.48550/arxiv.2010.12715}
Jagadeesh Balam, Jocelyn Huang, Vitaly Lavrukhin, Slyne Deng, Somshubra
  Majumdar, and Boris Ginsburg,
\newblock ``Improving noise robustness of an end-to-end neural model for
  automatic speech recognition,'' 2020.

\bibitem{ressl}
Mingkai Zheng, Shan You, Fei Wang, Chen Qian, Changshui Zhang, Xiaogang Wang,
  and Chang Xu,
\newblock ``Ressl: Relational self-supervised learning with weak
  augmentation,'' 2021.

\bibitem{paul1992design}
Douglas~B Paul and Janet Baker,
\newblock ``The design for the wall street journal-based csr corpus,''
\newblock in {\em Speech and Natural Language: Proceedings of a Workshop Held
  at Harriman, 1992}.

\end{thebibliography}

\end{document}